# [1]Some remarks on resolving and interpretation of the short marine magnetic anomalies.

S. A. Ivanov[1] and S. A. Merkuryev[1,2]

[1]*Pushkov Institute of Terrestrial Magnetism of the Russian Academy of Sciences, St Petersburg Filial. 1 MendeleevskayaLiniya, St Petersburg* 199034, *Russia, email: sergei.a.ivanov@mail.ru*
[2]*Saint Petersburg State University, Institute of Earth Sciences,Universitetskayanab.,7-9, St. Petersburg* 199034, *Russia*

**Abstract**. Marine magnetic anomalies of the tiny wiggles (TW) type can be used to solve geohistorical andpaleomagnetic problems. The model fields corresponding to Paleocene-Eocene anomalies in the northwestern Indian Ocean, which were formed during the fastspreading stage, were studied. For these fields, widelyused interpretation methods were compared. The testingwas performed with first the classical block model and then more complex models reflecting actual processesof oceanic accretion and magnetic field variations in the past.Spectral and statistical methods are used to estimate the magnetic anomalies(MA) resolving.Preprocessing of a set profiles based on the maximal correlation is considered.

**Keywords:** marine magnetic anomalies, geochronological analysis, spreading model, magnetic modeling.

## Introduction

Progress in studying the timescale fine structure isprimarily related to studies of MA at fastspreading centers, where linear anomalies with smallamplitudes (25–100 nT) and short periods weredetected among typical largeamplitude anomalies(Blakely and Cox, 1971; Emilia and Heinrichs, 1972;Cande and LaBrecque, 1974; Bouligand et al., 2006).These anomalies, called tiny wiggles in (LaBrecqueetal., 1977), are studied and discussed because it hasnot yet been determined whether these anomalies represent unknown short polarity chrons or correspond topaleomagnetic field intensity fluctuations. The unknownorigin of TW MAresulted in theappearance of the cryptochron special term in magnetostratigraphy, which is used to denote globallymapped geomagnetic singularities shorter than 30 kyr(Cande and Kent, 1992a, 1992b). In addition to thepaleomagnetic aspect of studies, TWtype anomaliescan be used to specify a spreading rate variation.Therefore, it becomes more important to determine amethod that can be used to study anomalies, the possible errors in information about variations in the ancient magnetic field.

## Data and Method

We consider the modeling methods for the Vine-Matthews (VM) classical model and more complexmodels reflecting actual processes of oceanic accretion and magnetic field variation in the past. According to the **classical two-dimensional block spreading model** by Vine-Matthewsit is actually assumed that magnetization *j(x)*reversed instantaneously in time and that polarity blocks have vertical faces. In this case theoretical magnetic anomalies m(x) results from the

---

[1]The research was supported in part by theRussia Foundation of Basic Research grants 15-05-06292.

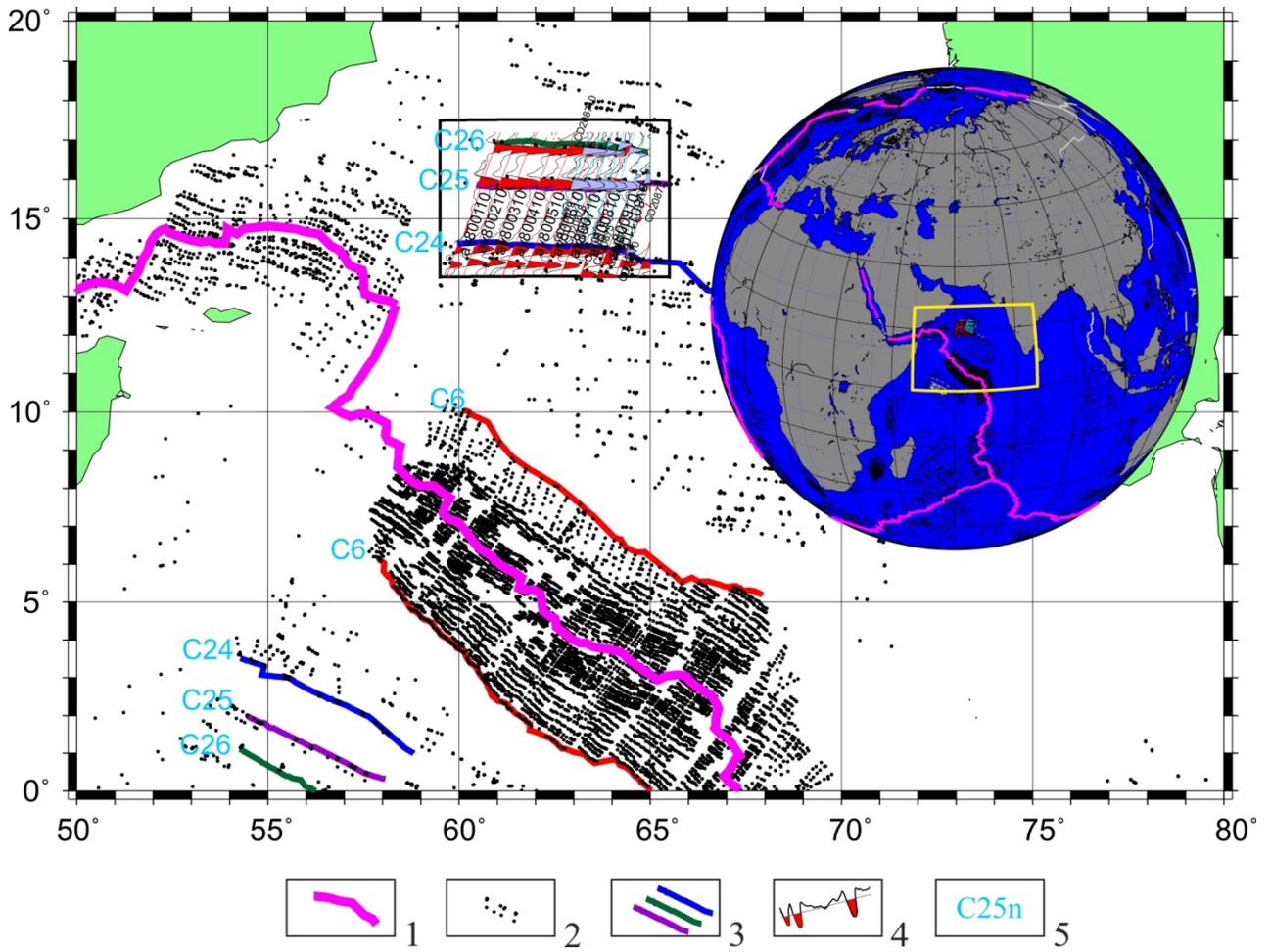

Fig.1.Map of linear magnetic anomalies in the northwestern Indian Ocean according to (Seton et al., 2014) and magnetic profiles, crossing anomalies A24 A25and A26: the plate boundary (1), crossings of MA (2), axes of linear magnetic anomalies (3), magnetic profile (4), and polarity chron denotation(5).Yellow box in the inset map outlines the region shown by the larger map. Magnetic anomalies in the areas outlined in black box are shown in Fig.3.

convolution$j(x)$ and $f_E(x)$ where $f_E(x)$ is a function that depends on depths of the upper (a) and lower (b) edges and the magnetization vector orientation. In the Fourier domain, the convolution procedure is equivalent to a simple multiplication of Fourier images:

$$M(\omega) = J(\omega) \cdot F_E(\omega).$$

As is shown in (Schouten and McCamy, 1972), the MA Fourier transform (spectrum) can be written in the following form for an ideal spreading model:

$$M(\omega) = J(\omega) \cdot F_E(\omega) = J(\omega)2\pi C e^{-i\theta}\left(e^{-|\omega|a} - e^{-|\omega|b}\right),$$

where $\theta$ is the parameter characterizing the anomaly asymmetry (skewness) and C is the amplitude coefficient. This expression indicates that MAobserved at a certain distance from the magnetic layer fundamentally represent initial signal $J(\omega)$ which came through a bandpass filter, $F_E(\omega) = 2\pi C e^{-i\theta}\left(e^{-|\omega|a} - e^{-|\omega|b}\right)$, i.e., an Earth filter (Schouten, 1971; Schouten and McCamy,

1972).For usual seafloor parameters a = 3.0 km and b = 3.5 km the bandpass is 10–50 km, thus, shortperiod anomalies of the TW type with periods shorter than 10 km are "filtered" with an Earth filter, since the magnetic layer is located far from the observation level.

For more realistic oceanic crust structure the magnetic anomaly spectrum can generally be written in the following form:

$$M(\omega) = J(\omega) \cdot F_E(\omega) \cdot F_R(\omega).$$

The eruption of new lavas (the extrusive process (Schouten and McCamy, 1972)) and the formation of vertical dikes (the intrusive process (Harrison,1968) are often modeled with Gaussian functions g(x) with the parameter σ. The spectral density of this function is well known:

$$F_R(\omega) = \exp(-2k^2\sigma^2)$$

We have estimated resolvability of marine magnetic anomalies using the two approaches.

**Estimation Based on a Spectral Approach.**

Analysis of the spectrum makes it possible to estimate the source depth. If deep sources exist, the contribution of low frequencies increases and distributed stochastically. If also the sources are distributed stochastically, the slope (α) of the energy spectrum natural logarithm plot is related to the source depth (z) by $\alpha = \arctan(2z)$. If the energy spectrum slopes insignificantly for high frequencies, this indicates that the sources are located near the ocean surface and are independent of the basaltic layer (Spector and Grant, 1970). The shortest obtained wavelength above the noise level is approximately 3–5 km. This means that the position of blocks generating anomalies, which are located at a distance of 3–5 km and more from one another, can theoretically be determined.

**Estimation based on the Cramer-Rao inequality**

We apply the Cramer-Rao method of mathematical statistics in order to estimate the resolvability of small amplitude short period anomalies, see details in (Ivanov and Merkuryev, 2013). We first consider the VM model. Theoretical MA calculated from the magnetic layer were constructed with the use of the Cande and Kent (1995) scale at a spreading rate of 6.7 cm yr$^{-1}$ with a quantization interval of 1 km. Normally distributed noise with an rms deviation of σ = 3 nT is imposed on the field. Note that the obtained minimal error values vary proportionally at different values of σ. Both for VM model and the smoothed model the block width is determined much more accurately than the block center. Specifically, for a block with a width of 264 m, the minimal rms errors are 13 and 140 m when the block width and center are determined, respectively. It was also shown that the minimal error of the block boundary determination do not depend on information about the position of adjacent blocks. For the VM model smoothed by convolution with the Gaussian function at σ=2 the Cramer-Rao estimates of the minimal error increase by a factor of 1.5–2.

## Interpretation of TW Model Magnetic Anomalies

Using a model example, we consider the application of some popular methods developed in order todetermine magnetization and/or polarity reversalboundaries. Figure 2 shows theoretical MAand the results of their inversions. The magnetic layer structure in the VM classical spreadingmodel with vertical boundaries of blocks of oppositepolarities is present in Fig. 2 (bottom). The top part ofFig. 2 illustrates the inversions of theoretical anomalies performed by widely known methods: the analytical signalNabighian M.N.), the magnetization calculation by theParker-Huestismethod (Parker-Huestis, 1974), andthe analytical downward continuation. A comparison ofthe inversion results with the initial magnetization distribution indicates that all of these methods make it possibleto judge only the position of large block boundaries; thefine structure of the field cannot be recovered.

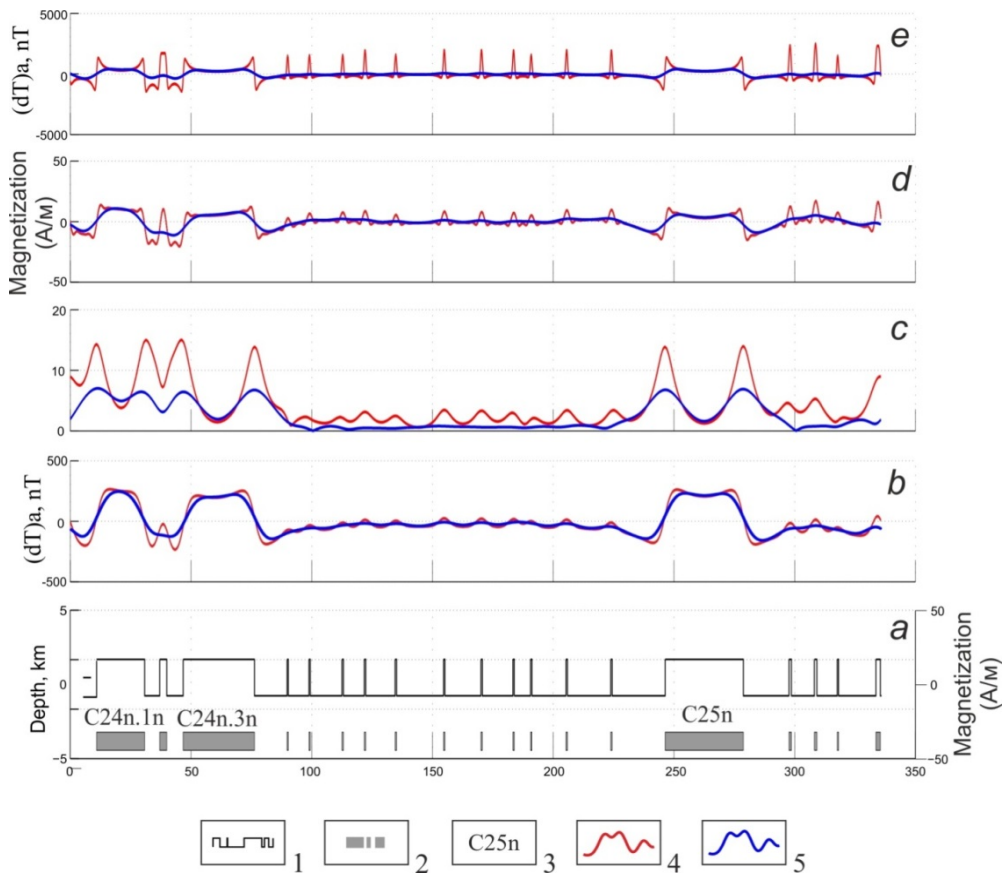

Fig.2.Methods for interpreting marine magnetic anomalies compared with the use of a model example for the classical andsmoothed spreading models. (a) The magnetic layer; (b) magnetic anomalies; (c) the analytical signal; (d) effective magnetizationcalculated using the Parker and Huestis (1974) method; (e) the analytical downward continuation of the anomalous magneticfield: magnetization (1), the magnetic layer in the VM classical model with blocks of direct polarity colored gray (2), polaritychron denotation (3), and magnetic field for the classical (4) and smoothed (5) models and their transformations.

**Preprocessing of profiles.**

One of the ways to improve TW interpretation is to use a stack. Let we have a set of profiles in the same region.The profiles are digitized with the same sampling rate but without reference to the middle ocean ridge. To solve themagnetic inverse problem using the stack we need toalign the profiles. We propose a procedure which can perform this automatically.

First, expand all profiles by zero before and after the sampling data. Apply the random multistart method (the Monte Carlo method), exactly, shift randomly part of profiles and as a pre-stack take the profile, which is at any point equal to the mean of the shifted profiles. Find the correlation coefficients of this pre-stack and find the sum $C$ of these coefficients. Starting from the new set of profiles shift again several profiles choosethe configuration with the larger $C$. Repeat these iterations and fix the best configuration. Then take again the original profiles and repeat all steps. Find the shifts when $C$arrives its maximal value what gives the stack.

We have used the original software based on algorithm described above to align theseveral magnetic profiles from the NW part Indian ocean crossed anomalies 24, 25 and 26. Fig.3 shows theoriginal profiles, the profiles after the preprocessing and the stacked profile compared with spreading model.The stacked profile (Figure 3, c) reveals several small scale anomalies (TW) betweenanomalies 24, 25 and 26 that may be indicativeof short polarity intervals.Synthetic magnetic anomalies computed for oceanic crust created between 58 and 52 Ma (using geomagnetic reversal timescale of Candy and Kent [1995]at a spreading rate of 6.7 cm yr$^{-1}$) at 10°S along a ridge trend N90°E, observed at 15°N and 63°E. Our estimation shows, that alignement and averaging of the observed profiles can improve correlation between model profile and stack profile.

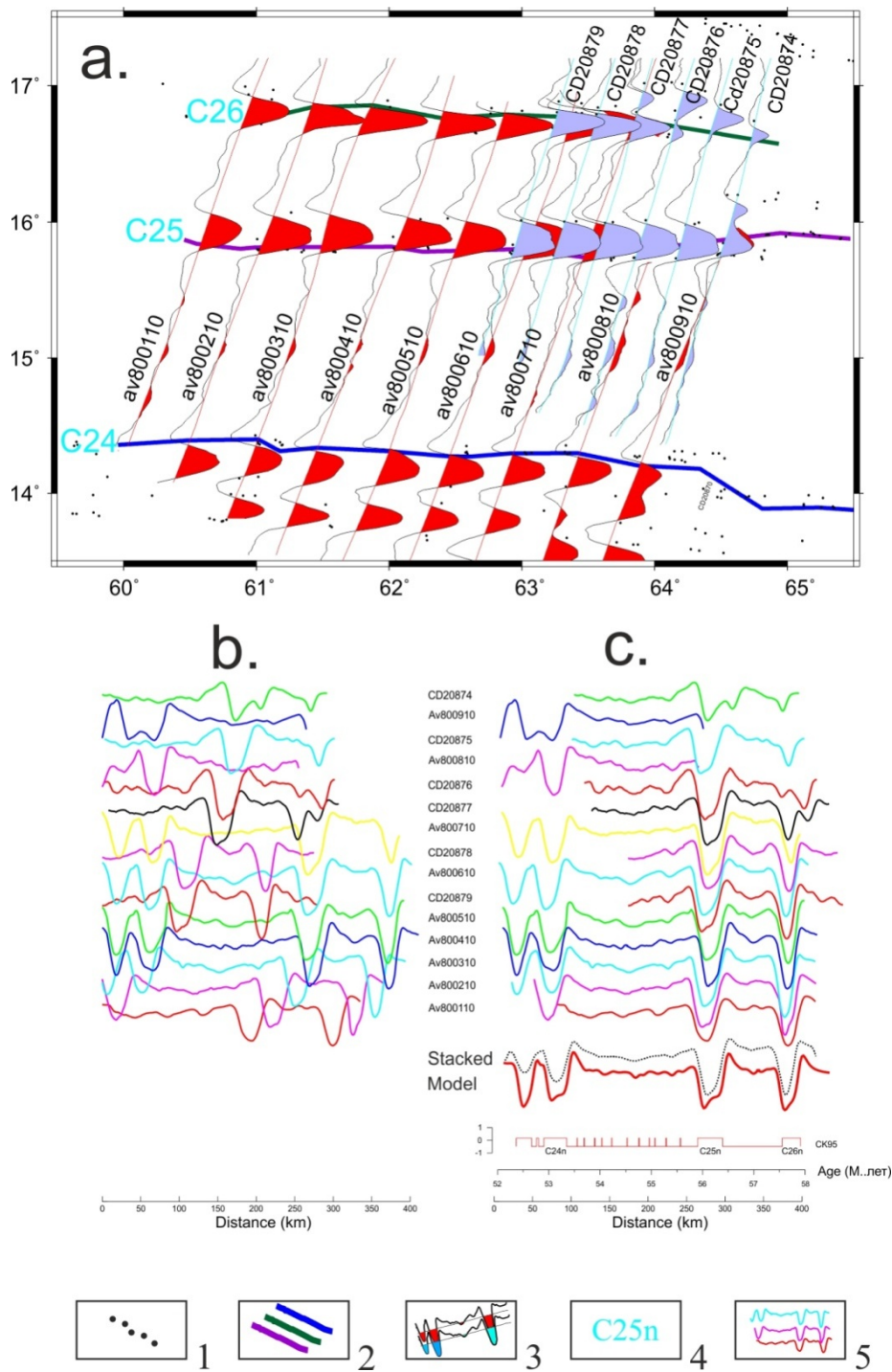

Fig.3 Observed magnetic anomalies profiles in the northwestern Indian Ocean(a,b), that were aligned using proposed technics and averaged to obtain the stack profile(c). Anomalies data acquired by R/V "Charles Darvin" and "Admiral Vladimirsky". Crossings of MA(1), axes of linear magnetic anomalies (2), magnetic profiles are projected orthogonally to ship tracks, positive to the east(3), polarity chron denotation(4), magnetic profiles before alignement. See caption to Fig.1 and Fig.2 for further information.

## Conclusions

We showed that the lateral structure of sources, i.e., the contact position, cannot be reconstructed when classical methods are used to solve the paleomagnetic problem. The estimation of the contact position minimal determination errors for the VM classical model and a more realistic model are obtained. An approach to align the set of profiles is considered.